\documentclass[lettersize,journal]{IEEEtran}
\usepackage{amssymb}
\usepackage{amsthm}
\usepackage[utf8]{inputenc}
\usepackage{amsmath}
\usepackage{graphicx} 
\usepackage{multirow}
\usepackage{graphics}
\usepackage{graphicx}
\usepackage{subfigure}
\usepackage{array}
\usepackage{booktabs}
\usepackage{cite}
\usepackage[colorlinks]{hyperref}
\usepackage[table,xcdraw]{xcolor}
\usepackage[ruled, linesnumbered]{algorithm2e}
\usepackage{makecell}
\usepackage{tikz,xcolor,hyperref}
\definecolor{lime}{HTML}{A6CE39}
\DeclareRobustCommand{\orcidicon}{%
    \begin{tikzpicture}
    \draw[lime, fill=lime] (0,0) 
    circle [radius=0.16] 
    node[white] {{\fontfamily{qag}\selectfont \tiny ID}};    \draw[white, fill=white] (-0.0625,0.095) 
    circle [radius=0.007];    \end{tikzpicture}
    \hspace{-2mm}}
\foreach \x in {A, ..., Z}{%
    \expandafter\xdef\csname orcid\x\endcsname{\noexpand\href{https://orcid.org/\csname orcidauthor\x\endcsname}{\noexpand\orcidicon}}
    }

\begin{document}

\title{Multi-Domain Virtual Network Embedding Algorithm based on Horizontal Federated Learning}

\author{
Peiying Zhang\orcidB{}, ~\IEEEmembership{Member,~IEEE}, Ning Chen\orcidD{}, ~\IEEEmembership{Student Member,~IEEE}, Shibao Li\orcidC{}, \\Kim-Kwang Raymond Choo\orcidG{}, ~\IEEEmembership{Senior Member,~IEEE,} and Chunxiao Jiang\orcidE{}, ~\IEEEmembership{Senior Member,~IEEE}

\thanks{
    This work is partially supported by the Shandong Provincial Natural Science Foundation, China under Grant ZR2020MF006, partially supported by the Industry-university Research Innovation Foundation of Ministry of Education of China under Grant 2021FNA01001, partially supported by the Major Scientific and Technological Projects of CNPC under Grant ZD2019-183-006, partially supported by the Open Foundation of State Key Laboratory of Integrated Services Networks (Xidian University) under Grant ISN23-09. The work of K.-K. R. Choo was supported only by the Cloud Technology Endowed Professorship.
    (\textit{Corresponding authors:} \textit{Peiying Zhang}).
    }
\thanks{Peiying Zhang and Ning Chen are co-first authors of this work.}
\thanks{Peiying Zhang and Ning Chen are with the College of Computer Science and Technology, China University of Petroleum (East China), Qingdao 266580, China, and also with the State Key Laboratory of Integrated Services Networks, Xidian University, Xi'an 710071, China. (\textit{email: zhangpeiying@upc.edu.cn, nchen@s.upc.edu.cn})}
\thanks{Shibao Li is with the College of Oceanography and Space Informatics, China University of Petroleum (East China), Qingdao 266580, China. (\textit{email: Lishibao@upc.edu.cn})}
\thanks{Kim-Kwang Raymond Choo is with the Department of Information Systems
and Cyber Security, University of Texas at San Antonio, San Antonio, TX
78249-0631 USA. (\textit{email: raymond.choo@fulbrightmail.org})
}
\thanks{Chunxiao Jiang is with the National Research Center for Information Science and Technology, Tsinghua University and the Tsinghua Space Center, Beijing 100084, China (e-mail: jchx@tsinghua.edu.cn).}
}

\markboth{IEEE Transactions on Information Forensics \& Security}%
{Zhang \MakeLowercase{\textit{et al.}}: Multi-Domain Virtual Network Embedding Algorithm based on Horizontal Federated Learning}


\maketitle

\begin{abstract}
Network Virtualization (NV) is an emerging network dynamic planning technique to overcome network rigidity. As its necessary challenge, Virtual Network Embedding (VNE) enhances the scalability and flexibility of the network by decoupling the resources and services of the underlying physical network. For the future multi-domain physical network modeling with the characteristics of dynamics, heterogeneity, privacy, and real-time, the existing related works perform unsatisfactorily.
Federated learning (FL) jointly optimizes the network by sharing parameters among multiple parties and is widely employed in disputes over data privacy and data silos. Aiming at the NV challenge of multi-domain physical networks, this work is the first to propose using FL to model VNE, and presents a VNE architecture based on Horizontal Federated Learning (HFL) (HFL-VNE).
Specifically, combined with the distributed training paradigm of FL, we deploy local servers in each physical domain, which can effectively focus on local features and reduce resource fragmentation. A global server is deployed to aggregate and share training parameters, which enhances local data privacy and significantly improves learning efficiency. Furthermore, we deploy the Deep Reinforcement Learning (DRL) model in each server to dynamically adjust and optimize the resource allocation of the multi-domain physical network. In DRL-assisted FL, HFL-VNE jointly optimizes decision-making through specific local and federated reward mechanisms and loss functions. Finally, the superiority of HFL-VNE is proved by combining simulation experiments and comparing it with related works.
\end{abstract}


\section{Introduction}
\IEEEPARstart{N}{etwork} virtualization (NV) is an emerging technology for the dynamic planning of Internet architecture \cite{9651548, 9372899}. By slicing the network, it can effectively overcome the rigidity of the network architecture. Specifically, by sharing infrastructure, NV enables multiple heterogeneous virtual networks (VNs) deploy in the same physical network, thereby meeting users' diverse needs for network services \cite{wang2021distributed}. Furthermore, since infrastructure resources are shared by multiple virtual networks, this significantly improves the utilization of physical network resources and avoids inefficient utilization of infrastructure. Therefore, the key challenge in realizing network virtualization is how to efficiently assign physical network resources to VNs, i.e., to efficiently embed virtual networks into physical networks, which is defined as the virtual network embedding (VNE) problem \cite{9475485}.

\begin{figure*}
    \centering
    \subfigure[Centralized Training Paradigm]{
        \includegraphics[width=0.48\linewidth]{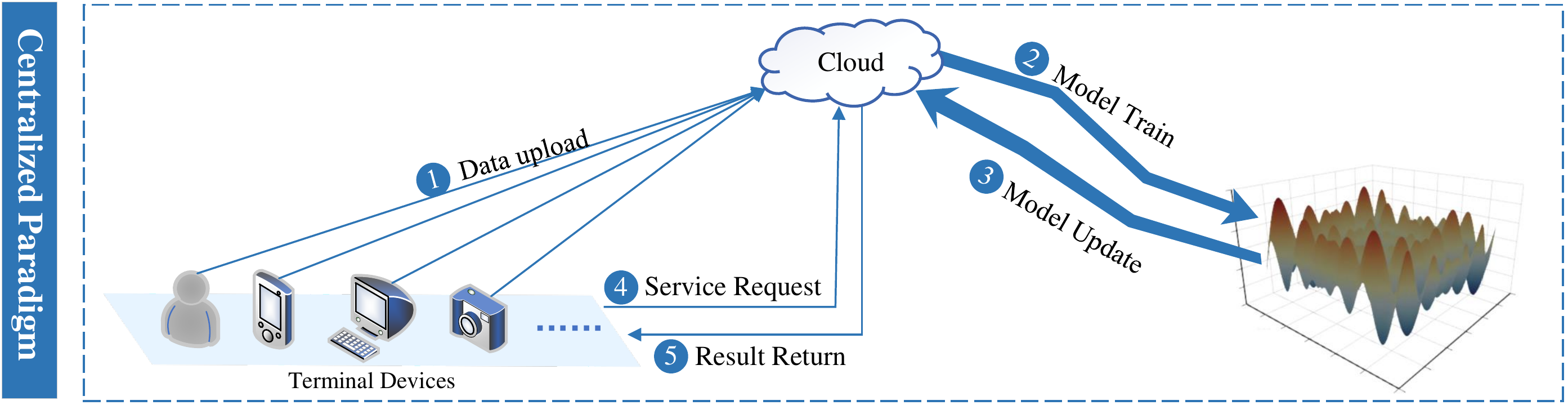}
        \label{fig: back1}
    }
    \subfigure[Federal Learning Training Paradigm]{
        \includegraphics[width=0.48\linewidth]{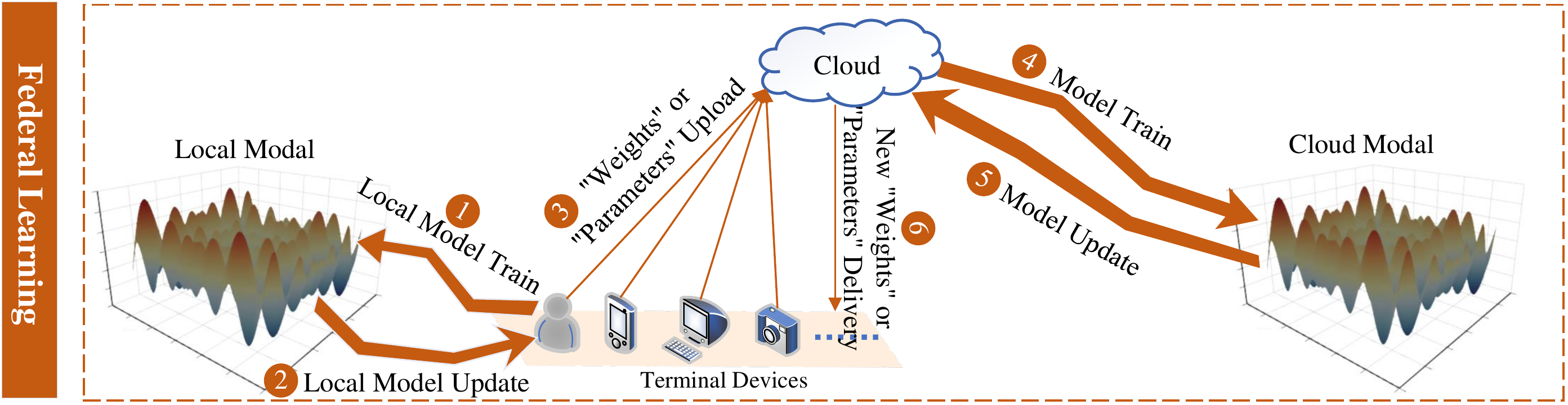}
        \label{fig: back2}
    }
    \caption{Traditional centralized training paradigm and federated learning training paradigm}
    \label{fig:1}
\end{figure*}
VNE improves the scalability of the architecture by decoupling the underlying physical network resources and services \cite{yao2020continuous}. Specifically, Infrastructure Providers (InPs) and Internet Service Providers (ISPs) are the principal managers of the network architecture. The former is responsible for the deployment and maintenance of network equipment, and the latter is responsible for deploying network protocols and providing end-to-end services. The ISP proposes a virtual network request (VNR) for the corresponding network service, and the InP allocates corresponding physical resources to it according to the VNR. It should be noted that such system optimization problems can be effectively modeled as Markov Decision Processes (MDPs) \cite{9653840}. Furthermore, VNE is notably an NP-hard problem. Therefore, the key to VNE optimization is to ensure the quality of service (QoS), while responding to VNR as much as possible, improving the revenues of embedding decisions, and reducing energy consumption \cite{zhang2019virtual}.

The traditional static VNE algorithm is based on the premise that infrastructure and resource requests remain unchanged. However, it is not applicable in the dynamic reality and future network \cite{yan2020automatic}. Since the existence of the life cycle of the VNR, it makes the resource allocation mode change in real-time.
And, variations in user requirements will lead to variations in VN topology, scale, and resource requirements.
Furthermore, in the long run, the physical network will exhibit time-varying characteristics\cite{9505612}. Specifically, the resource distribution of the physical network, as well as the network topology, changes over time.
Especially in multi-domain physical networks, the existence of the above problems leads to the key difficulty of VNE modeling. 
In addition, resources occupied by new VNEs after being successfully embedded and resources released by VNRs at the end of their life cycles will lead to fragmentation of physical network resources over time\cite{9475485}. It will in turn lead to fragmentation of the embedding process, lower VNR acceptance rates, and lower embedded revenue.


In addition, under the rapid development of the new generation of communication infrastructure, related technologies, e.g., edge computing and Internet of Things (IoT), will promote the generation of large amounts of data by network devices \cite{9460016}. The traditional VNE algorithm adopts a centralized training paradigm, as shown in Fig. \ref{fig: back1}, which uploads all data to the cloud server, and the model deployed in the server trains all the data and updates the model \cite{zhang2019virtual}. In the final application, the cloud service needs to be requested locally. However, this paradigm has shortcomings such as weak data privacy, poor real-time performance, high central pressure, and long training time \cite{yao2020continuous}.
In conclusion, the VNE algorithm still needs to be further improved in the dynamic, heterogeneous, private, real-time, limited hardware, and other network environments.


As a distributed ML paradigm, Federated Learning (FL) is originally proffered to solve the privacy problem of user-uploaded data and has been further used to solve the phenomenon of data silos between enterprises \cite{9690152}. The learning process of FL is shown in Fig. \ref{fig: back2}, each participant can conduct joint modeling with the help of other parties' data.
FL deploys ML algorithms on different terminals in a distributed manner, and establishes a shared ML model through multi-party cooperation for global macro control \cite{mothukuri2021federated}.
It can significantly concentrate the local and global dynamics and heterogeneity, and effectively diminish the fragmentation of physical resources in the VNE algorithm \cite{YANG202233}. Without sharing data, the local model builds shared ML by uploading encrypted parameters or weights to control the global, which makes data transmission highly private and reduces storage and computational pressure \cite{9460016}. In addition, different from the previous centralized training paradigm, FL significantly reduces the model learning time and the response time of demand through the joint optimization of local training and cloud-shared models \cite{zhang2021bc}.


Horizontal Federated Learning (HFL) can be applied to environment where the datasets of each participant in FL have the same feature space and different sample spaces \cite{9551794,9735274}. According to the VNE problem of multi-domain networks, the nodes and links of each physical domain have the same feature space, so we use HFL to model the multi-domain VNE. Therefore, based on the advantages of FL, we propose a novel virtual network embedding algorithm based on horizontal federated learning (HFL-VNE), which can effectively relieve the shortcomings of traditional static and centralized training paradigm VNE algorithms. To sum up, the main contributions and innovations are:
\begin{itemize}
    \item Aiming at the shortcomings of existing VNE algorithms in modeling the dynamics, heterogeneity, privacy, and real-time performance of multi-domain physical networks, this work proposes a VNE architecture based on HFL. To our knowledge, this is the first work utilizing FL to drive resource allocation.
    \item To better fit the dynamics and heterogeneity of physical networks, we combine the distributed training paradigm in FL and deploy local servers in each physical domain. It focus on the characteristics of each local domain, thereby effectively reducing the phenomenon of resource fragmentation. In addition, for privacy and real-time issues, we deploy a global server on the overall network architecture. On the premise of not sharing local data, the global server aggregates the shared weights or parameters from each local server, which not only strengthens the confidentiality of the data but also reduces the central pressure, thereby improving the learning efficiency.
    \item To promote the VNE modeling better interact with the environment, we deploy a similar DRL model in each server of FL to dynamically adjust and optimize the resource allocation of multi-domain physical networks. In particular, the local DRL model is responsible for calculating and screening candidate physical nodes that satisfy the established relevant constraints. The global DRL model combines the federated state and federated action to control the global macroscopically, and feedback federated reward to each physical domain.
    \item Eventually, the superiority and effectiveness of the proposed HFL-VNE algorithm are comprehensively verified on multiple evaluation indicators through a simulation case study.
\end{itemize}

The remainder of this work is in turn: Section \ref{sec:2} analyzes related works. Section \ref{sec:3} introduces the definition and modeling of the problem. Section \ref{sec:4} lists the relevant indicators. Section \ref{sec:5} documents the design and implementation details of HFL-VNE. Section \ref{sec:6} presents the simulation case setup and comparative analysis of the results. Eventually, Section \ref{sec:7} summarizes our done.
\section{Related Work}\label{sec:2}
Existing VNE works can be mainly divided into two categories: heuristic search strategies and artificial intelligence (AI)-based strategies.
\subsection{Heuristic Search Strategies}
Heuristic algorithms adjust decisions through statistical analysis, numerical optimization, etc \cite{8122008}. The classic algorithms are D-ViNE and R-ViNE proposed by Chowdhury et al. \cite{5061987}, they specify the embedding process of VN as a mixed-integer procedure based on deterministic and random rounding techniques, respectively. In addition, its research proves that the ultimate goal of the VNE strategy is to reduce resource cost and increase resource revenue and the VNR acceptance rate.
Furthermore, Cheng et al. \cite{cheng2011virtual} propose a VNE algorithm based on topology-aware node ranking (NodeRank). Specifically, it rank based on the resources of physical nodes as well as the network topology, and rank according to the Markov random walk model. The embedding process adopts two strategies,
which are based on virtual node level embedding and breadth-first search (BFS).

In addition, including our previous related work, considering that the previous works have not considered the constraints of storage resources, we propose a 3-dimensional resource embedding strategy in accordance with computing, storage, and network. And two VNE strategies are designed respectively founded on two-stage mapping, NRM-VNE, and RCR-VNE \cite{7976281}. Moreover, inspired by the introduction of local search in the meta-heuristic algorithm, we also propose a VNE strategy based on an improved genetic algorithm \cite{zhang2019virtual}, which significantly improves various evaluation indicators.

However, such heuristic search strategies are difficult to exhibit satisfactory performance in today's complex and highly dynamic physical networks. Subjectively specified embedding rules limit the flexibility and scalability of allocation strategies \cite{8672596}. Furthermore, most methods adopt a static VNE strategy, which becomes less practical in real-time changing VNR as well as in physical networks.

\subsection{AI-based Strategies}
With the maturity of AI technology, it has been widely used in network modeling and has significantly enhanced the intelligence, generalization, and accuracy of the network\cite{9454579, DUAN2022108676}.
To dynamically focus on the physical network environment, Yao et al. propose a reinforcement learning (RL)-based embedding algorithm (CDRL) \cite{yao2020continuous}, which treats the embedding process of physical nodes as a time-series problem and proposes a strategy for continuous embedding. Comparative results demonstrate that CDRL possesses a significant advancement in various indicators.
Combining the powerful perception capability of deep learning (DL), our previous work proposes dynamic VNE strategies based on graph convolutional networks (GCN) and RL (GCN-RL) \cite{9475485}. Based on the fitness matrix, it focuses on the physical components with high fitness for VN embedding. In addition, GCN can better pay attention to the topology of the network and extract network environment features in real-time. The final experiments show that it significantly reduces the fragmentation of network resources.

RL has good decision-making ability but poor perceptual performance. DL, on the other hand, has the opposite properties of RL. The emerging Deep Reinforcement Learning (DRL) \cite{9403369} in the field of AI, combining DL and RL at the same time makes it applicable to potential problem areas. DRL effectively perceives the environment through the deep neural network (DNN), which combines the efficient interaction among the agent and physical network and the positive feedback of the reward mechanism, so that the model has excellent perception and decision-making capabilities at the same time \cite{wurman2022outracing}. Yan et al. \cite{yan2020automatic} propose an automatic embedding VNE algorithm using GCN as an intelligent agent for DRL. Experiments show that it has better robustness and performance of various indicators. Furthermore, our previous work \cite{9505612} combined with Convolutional Neural Network (CNN) proposes an intelligent agent with a 5-layer architecture for computing physical node embedding probabilities and screening candidate physical nodes.

However, most of these related works adopt a centralized training paradigm, which deploys only one ensemble model globally responsible for the global VN embedding. There is no specific focus on the local characteristics of each physical domain, and there is room for improvement in reducing resource fragmentation. In addition, in these works, the data of each physical domain is globally visible, and the data privacy, real-time training and model pressure are not considered. According to various advantages of FL, we combine FL to adopt distributed strategy and deploy DRL model to jointly optimize the VNE process.

\section{Problem Definition and Modeling}\label{sec:3}
\subsection{Definition and Modeling of Virtual Network Embedding}
\begin{figure}
    \centering
    \includegraphics[width=0.9\linewidth]{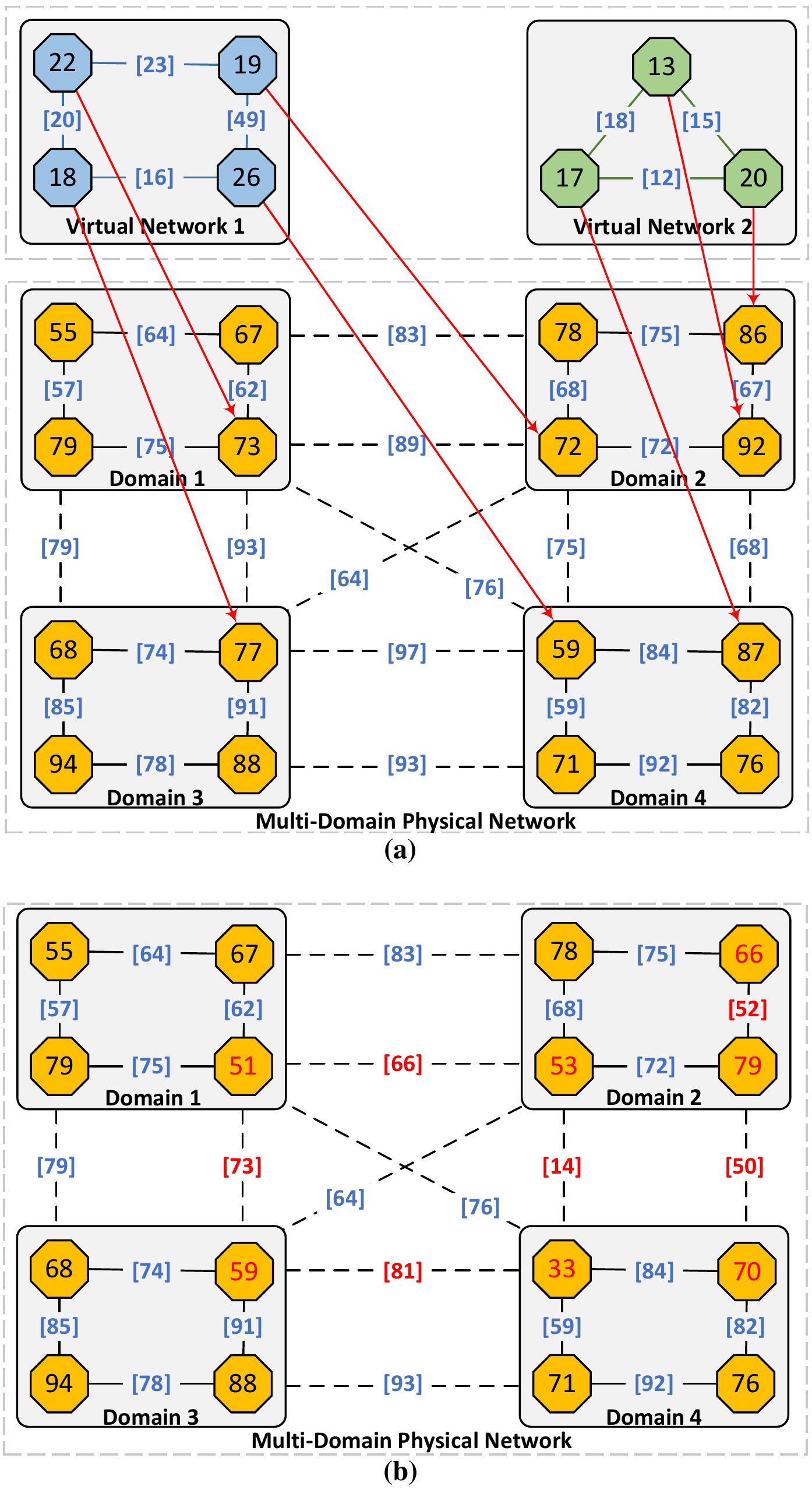}
    \caption{Schematic diagram of multi-domain VNE. (a) In the embedding process, the physical network arranges resources according to the resource requirements of the VNR. (b) After embedding, the physical network allocates resources to the virtual network.}
    \label{fig:vne}
\end{figure}
\begin{table}[]
\caption{The relevant symbols used in this study}
\label{tab:1}
\centering
\renewcommand\arraystretch{1.3}
\begin{tabular}{|l|l|l|}
\hline
\textbf{Network}                             & \textbf{Symbols} & \textbf{Definition} \\ \hline
\multirow{6}{*}{Physical Network $P$} & $OL^P$               &  The inter-domain link        \\ 
                                    &     $R\_OL^P$          &  The resource of inter-domain links          \\ 
                                    & $N^{P_i}$          &  The intra-domain node          \\ 
                                    & $IL^{P_i}$               &The intra-domain link            \\ 
                                    & $R\_IN^{P_i}$              &  The resource of intra-domain nodes          \\ 
                                    & $R\_IL^{P_i}$              &  The resource of intra-domain links          \\ \hline
\multirow{4}{*}{Virtual Network $V_i$} & $N^{V_i}$         &    The virtual nodes        \\
                                    & $L^{V_i}$        &    The virtual links        \\ 
                                    & $R\_N^{V_i}$               &  The resource of virtual nodes          \\ 
                                    & $R\_L^{V_i}$             &    The resource of virtual links        \\ \hline
\end{tabular}

\end{table}
We show an example of multi-domain VNE in Fig. \ref{fig:vne}, where the numbers represent resource attributes. Specifically, the ISP proposes the VNR and specifies configuration attributes such as the requested resource, and the InP embeds the resource as reported by the VNR. Furthermore, different VNRs can share the resources of the same physical network. Therefore, the main purpose of VNE is how to allocate physical network resources as reasonably and efficiently as possible. In terms of quantitative expression, that is, in the face of successive VNRs, how to respond to VNRs as much as possible with the lowest resource consumption.


To study the multi-domain VNE problem, we model the network for intuitive analysis and derivation of the problem. In this study, the relevant symbols used are recorded in Table \ref{tab:1}. Specifically, we model the multi-domain physical networks with multiple weighted undirected graphs, as:
\begin{equation}
\begin{aligned}
    P &= \left\{P_i, OL^P, R\_OL^{P}\right\},\\
    s.t. \quad P_i&=\left\{N^{P_i}, IL^{P_i}, R\_IN^{P_i}, R\_IL^{P_i}\right\},
\end{aligned}
\end{equation}
where $P_i$ represents the \textit{i}-th physical network.
Furthermore, the virtual network generated by successive VNRs is similarly modeled as:
\begin{equation}
    V_i=\left\{N^{V_i}, L^{V_i}, R\_N^{V_i}, R\_L^{V_i}\right\},
\end{equation}
where $V_i$ represents the \textit{i}-th virtual network.

It should be noted that in this study, we employ computing (CPU) resources to represent the resource attributes of nodes, and bandwidth (BW) resources to represent the resource attributes of links. Therefore, the resources of $P$ and $V_i$ can in turn be represented as:
\begin{align}
    R\_IN^{P_i} &= \left\{icpu^{P_i}_1, icpu^{P_i}_2, \cdots, icpu^{P_i}_{|N^{P_i}|}\right\},\\
    R\_IL^{P_i} &= \left\{ibw^{P_i}_1, ibw^{P_i}_2, \cdots, ibw^{P_i}_{|IL^{P_i}|}\right\},\\
     R\_OL^{P} & = \left\{obw^{P_i}_1, obw^{P_i}_2, \cdots, obw^{P_i}_{|OL^{P}|}\right\},\\
      R\_N^{V_i} & = \left\{cpu^{V_i}_1, cpu^{V_i}_2, \cdots, cpu^{V_i}_{|N^{V_i}|}\right\},\\
     R\_L^{V_i} & = \left\{bw^{V_i}_1, bw^{V_i}_2, \cdots, bw^{V_i}_{|L^{V_i}|}\right\},
\end{align}
where $|N^{P_i}|$, $|IL^{P_i}|$ and $|OL^{P}|$ represent the number of physical nodes and links in the multi-domain physical network, respectively. $|N^{V_i}|$ and $|L^{V_i}|$ indicate the number of virtual nodes and links.

As mentioned before, the VNR has a life cycle, that is, a VNR has a request start time $t_s$ and a request end time $t_e$. Therefore, the $i$-th VNR is modeled as:
\begin{equation}
    VNR_i=\left\{V_i, t_s, t_e\right\},
\end{equation}
According to the above analysis, the virtual network embedding problem can be abstractly modeled as:
\begin{equation}
\begin{aligned}
{for}\  &i=1,2,\cdots,|VNR|:\  {do}\\
    &VNR_i\left\{V_i, t_s, t_e\right\}\rightarrow P_{sub},
\end{aligned}
\end{equation}
where $|VNR|$ represents the number of VNR, and $P_{sub}$ represents the subgraph of $P$.

\subsection{Definition and Modeling of Federated Learning}

\begin{figure}[!htbp]
    \centering
    \includegraphics[width=\linewidth,height=7cm]{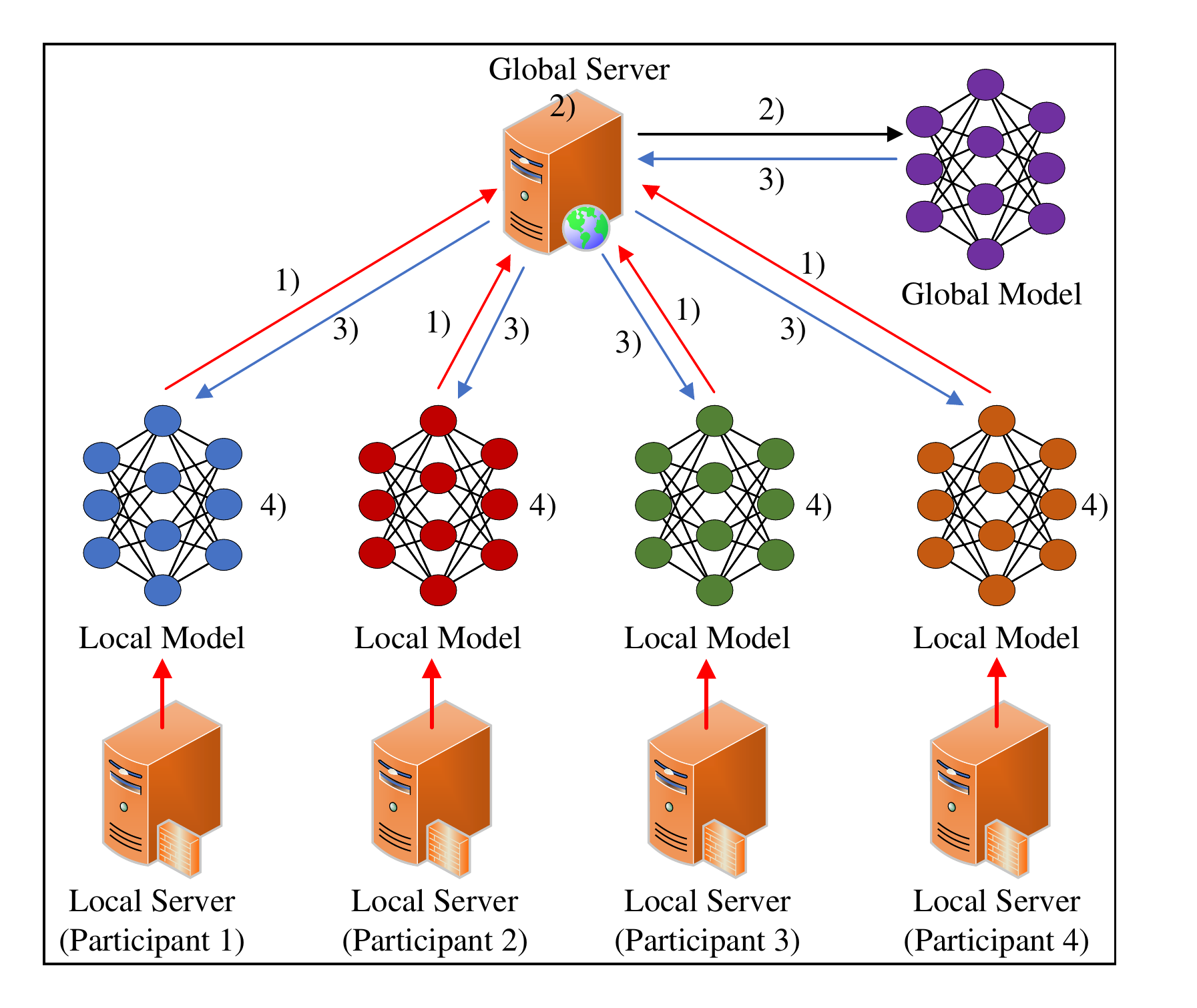}
    \caption{The general solution of HFL}
    \label{fig:hfl}
\end{figure}

The optimization goal of HFL is to ensure the privacy of information and accelerate distributed ML at the same time. Specifically, the local models in the local participant devices are trained using local data, and the model parameters and weights are uploaded to the central global aggregation server. The global model deployed in the global server is trained and optimized to perform macro-control on the global. Specifically, the general solution of HFL is show in Fig.\ref{fig:hfl}, as follows:
\begin{enumerate}
    \item Each participant trains a local model, minimizes the decision loss, and calculates the model gradient. Then, use encryption techniques such as homomorphic encryption to mask the weight and parameter information, and upload and update it to the global server.
    \item The global server securely aggregates the information transmitted by each local server, and updates the weights and parameters.
    \item The global model optimizes the training and sends the aggregated parameters and weight information to each participant.
    \item Each participant decrypts the received information, and uses the decrypted information to update their model parameters.
    \item Repeat 1)$\rightarrow$4) until the model converges.
\end{enumerate}

Furthermore, we assume that there are N physical domains. In each physical domain, we design a specific local ML model. Let the local dataset constructed by the environmental information in each physical domain $P_i$ be defined as $P_i=\left\{P_{i1}, P_{i2}, \cdots, P_{i|N_{P_i}|}\right\}$.
The construction process of the local ML model and the dataset will be described in the subsequent content. The local servers in each physical domain train their local models, and upload information such as weights and parameters. That information will be aggregated in the global server. It pdates and trains the weights and parameters, and sends the new parameters and weights $\theta$ to each local server. The local server downloads the information from the global server and updates the weights and parameters. Thus, each local server aggregates global information for training without using direct data information from other physical domains. It is also the superiority shown by the FL training paradigm in terms of security and privacy. Accordingly, the training objective of the local model of physical domain $P_i$ is to minimize its local loss $Local\_\mathcal{L}_i$:
\begin{equation}
    Local\_\mathcal{L}_i(\theta, P_i)=\frac{1}{|N_{P_i}|}\sum_{j=1}^{|N_{P_i}|}\mathcal{L}(\theta, P_{ij}),
    \label{eq:ll}
\end{equation}
where $\mathcal{L}(\theta, P_{ij})$ represents the loss value of sample $P_{ij}$ in the physical domain $P_i$. For the global model in the global server, its loss is defined as the weighted average of the local loss $ Local\_\mathcal{L}_i$ of each physical domain $P_i$. Therefore, its goal is to minimize the global loss $Global\_\mathcal{L}$:
\begin{equation}
\small
    Global\_\mathcal{L}(\theta, P)=\frac{1}{\sum_{i=1}^N|N_{P_i}|}
    \sum_{i=1}^N\left(|N_{P_i}|\cdot Local\_\mathcal{L}_i(\theta, P_i)\right).
    \label{eq:gl}
\end{equation}
\subsection{Adopted ML Model}
FL is generally regarded as a paradigm for collaborative learning of distributed ML models, therefore, the type of ML models deployed in FL is another key point for improving performance. 
To further enhance the presentation of our proposed HFL-VNE, we combine DRL to deploy local and global ML models to dynamically tune and optimize multi-domain physical networks. As an important goal-oriented technique, DRL optimizes the goal along a certain dimension through different steps. Specifically, the different components in DRL are expressed as follows:
\begin{figure}[!htbp]
    \centering
    \includegraphics[width=0.7\linewidth]{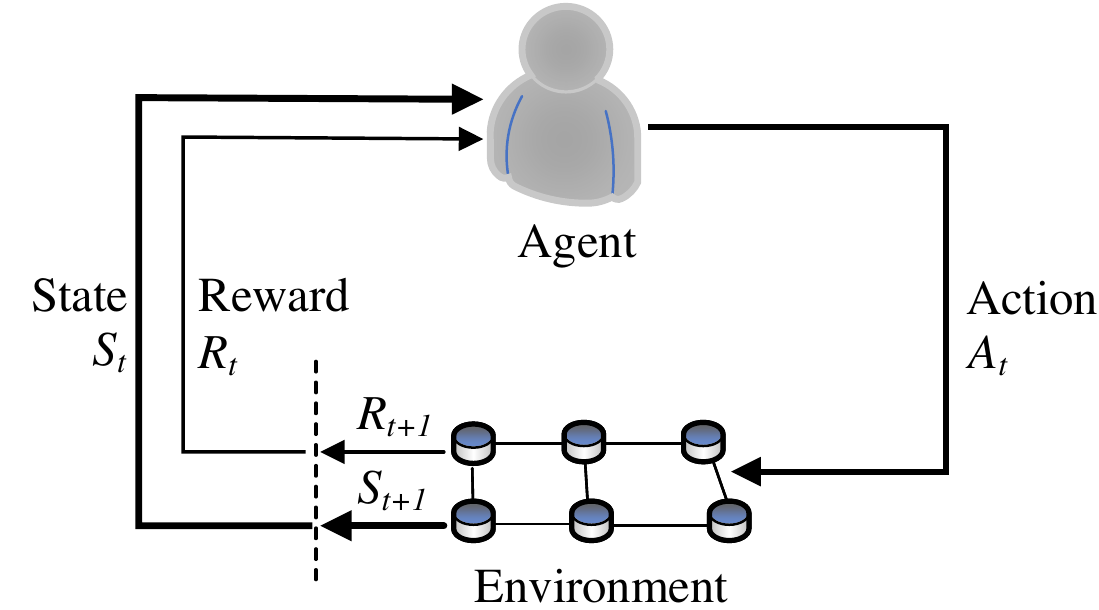}
    \caption{How the DRL model works}
    \label{fig:drl}
\end{figure}
\begin{itemize}
    \item Agent: It is mainly used as an important entity for the interaction between the model and the environment, and completes the interaction with the environment and makes corresponding decisions through the cooperation of the following components, as shown in Fig. \ref{fig:drl}.
    \item Action: It is mainly the decision-making adopted by the agent according to the environmental information.
    \item Reward: It is expressed as the feedback signal to the agent after the agent makes the corresponding action, which is mainly employed to evaluate the effectiveness of the current action. Specifically, if the positive feedback with a high reward is obtained, it means that the current action benefits the agent, and a similar action should be adopted in the next step. Conversely, the current action brings negative returns. Furthermore, it should be noted that the loss function optimized by the ML model in our proposed FL paradigm is consistent with the purpose of the reward mechanism here. Therefore, Eq.\ref{eq:ll} and Eq.\ref{eq:gl} are equivalent to the reward mechanism.
    \item State: It represents real-time information about the environment. It will serve as input to the agent so that the agent can interact with the environment in real-time. Furthermore, the variation of the state is often affected by actions taken by the agent.
\end{itemize}

\section{Related Indicators}\label{sec:4}
In this section, the relevant metrics used in the VNE problem will be introduced in detail.
\subsection{Basic Indicators}
We employ the character $\zeta_{n^{P}\rightarrow n^{V}}$ to denote whether the resources of the physical node $n^{P}$ are successfully allocated to the virtual node $n^{V}$, i.e., whether the $n^{V}$ is successfully embedded in the $n^{P}$.
\begin{equation}
   \zeta_{n^{P}\rightarrow n^{V}}=\begin{cases}
    1, & resource\  allocation\  succeeded,\\
    0, &  resource\  allocation\  failed,
    \end{cases}
\end{equation}
It should be noted that we use lowercase letters to indicate a specific element. For example, $n^{P}$ represents one of the nodes in the multi-domain physical networks.

Furthermore, we employ the character $\xi_{l^{P}\rightarrow l^{V}}$ to indicate whether the physical link $l^{P}$ are successfully allocated to virtual link $l^{V}$ of the virtual network, i.e., whether the $l^{V}$ is successfully embedded in the $l^{P}$.
\begin{equation}
   \xi_{l^{P}\rightarrow l^{V}}=\begin{cases}
    1, & resource\  allocation\  succeeded,\\
    0, &  resource\  allocation\  failed,
    \end{cases}
\end{equation}
It should be noted that, for convenience, we use $l^{P}$ to represent the collective term for $ol^p$ and $il^P$.

\subsection{Embedded Constraint Indicators}\label{sec:filter}
During the resource allocation process of the VNE, corresponding configuration constraints need to be followed. Only the resources of physical nodes or links that satisfy the relevant constraints can be successfully allocated. Specifically, they are:
\begin{align}
  &{\sum}_{j=1}^{|N^{V_i}|} \zeta_{n^{P}\rightarrow n^{V_i}_j}=1,\  \forall\  n^{V_i}_j\in N^{V_i},
    \label{eq:14}\\
   & {\sum}_{j=1}^{|L^{V_i}|} \xi_{l^{P}\rightarrow l^{V_i}_j}\geq 1,\  \forall\  l^{V_i}_j\in L^{V_i},
    \label{eq:15}\\
    &\begin{aligned}
 A\_N(n^P_k)&=R\_IN^P\left(n^P_k\right)\\&-\sum_{i=1}^{|VNR|}\sum_{j=1}^{|N^{V_i}|}\zeta_{n^{P_k}\rightarrow n^{V_i}_j}\times  R\_N^{V_i}(n^{V_i}_j),
\end{aligned}
\label{eq:16}\\
&A\_N(n^P_k)\geq 0,\  \forall \ n^P_k\in N^P,
    \label{eq:17}\\
    &R\_N^{V_i}(n^{V_i}_j)\leq A\_N(n^P_k),\  if \  \zeta_{n^{P}_k\rightarrow n^{V_i}_j}=1,
    \label{eq:18}\\
    &\begin{aligned}
 A\_L(l^P_k)&=R\_L^P\left(l^P_k\right)\\&-\sum_{i=1}^{|VNR|}\sum_{j=1}^{|E^{V_i}|}\xi_{l^{P}_k\rightarrow l^{V_i}_j}\times  R\_L^{V_i}(l^{V_i}_j),
\end{aligned}
\label{eq:19}\\
&A\_L(l^P_k)\geq 0, \  \forall \ l^P_k\in L^P,
    \label{eq:20}\\
 &   R\_L^{V_i}(l^{V_i}_j)\leq A\_L(l^P_k),\  if \  \xi_{l^{P}_k\rightarrow l^{V_i}_j}=1,
     \label{eq:21}
\end{align}
Where Eq. \ref{eq:14} indicates that the resources requested by each virtual node can only come from one physical node. Eq. \ref{eq:15} indicates that the resources requested by each virtual link can come from multiple physical links. Eq. \ref{eq:16} represents the available resources of a physical network node, which is expressed as the difference value between the physical node's initial resources and the allocated unavailable resources. Eq. \ref{eq:17} demonstrates the physical node available resources should be greater than 0. Eq. \ref{eq:18} indicates that the resources requested by the virtual node must smaller than the physical node's available resources that want to map. Eq. \ref{eq:19} demonstrates the physical link's available resources, which are expressed as the difference between the initial and the allocated resources. Eq. \ref{eq:20} demonstrates that the physical link available resources should be greater than 0. Eq. \ref{eq:21} indicates that the resources requested by the virtual link should be less than the physical link's available resources that you want to map.

\subsection{Evaluation Indicators}
The main optimization trend of the resource allocation strategy of VNE is to reduce resource consumption and improve resource benefit. The main purpose is to obtain the maximum resource allocation revenue with the least resource cost. Specifically, as many successful responses as possible should be made to the VNR. Therefore, to visually observe the performance, we use the following indicators to measure:
\begin{equation}
\footnotesize
\begin{aligned}
     R&(VNR_i)=R(V_i, t_s, t_e)\\&=\left(t_e-t_s\right)\left[\sum_{n^{V_i}\in N^{V_i}}R\_N^{V_i}(n^{V_i})+\sum_{l^{V_i}\in L^{V_i}}R\_L^{V_i}(l^{V_i})\right]\\
     &=\left(t_e-t_s\right)\left[\sum_{n^{V_i}\in N^{V_i}}cpu^{V_i}(n^{V_i})+\sum_{l^{V_i}\in L^{V_i}}bw^{V_i}(l^{V_i})\right],
\end{aligned}
\label{eq:22}
\end{equation}
\begin{equation}
\scriptsize
\begin{aligned}
     &C(VNR_i)=C(V_i, t_s, t_e)\\&=\left(t_e-t_s\right)\left[\sum_{n^{V_i}\in N^{V_i}}R\_N^{V_i}(n^{V_i})+\sum_{l^{V_i}\in L^{V_i}}R\_L^{V_i}(l^{V_i})hops(l^{V_i})\right]\\
     &=\left(t_e-t_s\right)\left[\sum_{n^{V_i}\in N^{V_i}}cpu^{V_i}(n^{V_i})+\sum_{l^{V_i}\in L^{V_i}}bw^{V_i}(l^{V_i})hops(l^{V_i})\right],
\end{aligned}
\label{eq:23}
\end{equation}
where Eq. \ref{eq:22} represents the revenue of a single VNE, and when the virtual network is allocated, the revenue will be generated. If the embedding process fails since the relevant embedding indicators are not met, no revenue will be generated. It is expressed as the sum of resources requested by the current VNR.
Eq. \ref{eq:23} represents the cost of a single VNE, and $hops$ represents the hops of the path. Intuitively, the cost is connected both the requested resource and the hops of the virtual path. The longer the path, the higher the link resource cost.

\begin{figure*}[!htbp]
    \centering
    \includegraphics[width=\linewidth, height=11 cm]{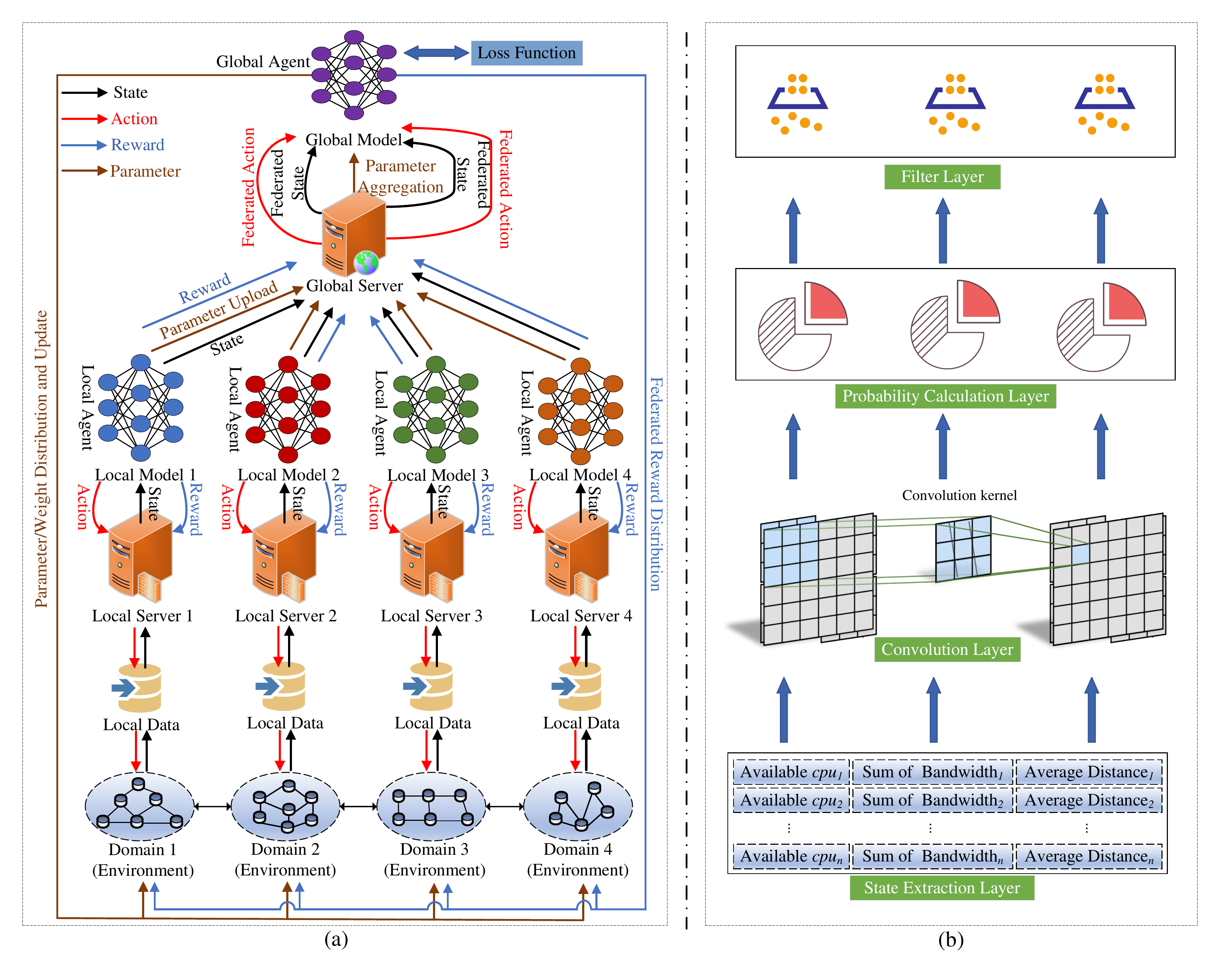}
    \caption{Architecture of HFL-VNE. (a) HFL architecture for multi-domain VNE with DRL as a learning-driven model. The local server deployed in the local physical domain is used for local precise control. The global model deployed by the global server is responsible for macro-control. DRL guides model training and efficient interaction with the environment through a reward mechanism. (b) The agent model of DRL in HFL-VNE consists of four layers: state extraction layer, convolution layer, probability calculation layer, and filter layer.}
    \label{fig:model}
\end{figure*}
Accordingly, the following indicators can be derived and utilized based on cost and revenue to evaluate VNE performance:
\begin{align}
    \begin{split}
       LtaR&=\lim_{T\rightarrow + \infty}\frac{\sum_{i=1}^{|VNR|}\int_{t=0}^T R(VNR_i)}{T}\\&=\lim_{T\rightarrow + \infty}\frac{\sum_{i=1}^{|VNR|}\sum^T_{t=0}R(VNR_i)}{T}, 
    \end{split}
\label{eq:24}\\
\begin{split}
     LtaR2C&=\lim_{T\rightarrow + \infty}\frac{\sum_{i=1}^{|VNR|}\int_{t=0}^T R(VNR_i)}{\sum_{i=1}^{|VNR|}\int_{t=0}^T C(VNR_i)}\\&=\lim_{T\rightarrow + \infty}\frac{\sum_{i=1}^{|VNR|}\sum^T_{t=0}R(VNR_i)}{\sum_{i=1}^{|VNR|}\sum^T_{t=0}C(VNR_i)},
\end{split}
\label{eq:25}\\
\begin{split}
     Acc&=\lim_{T\rightarrow + \infty}\frac{\int_{t=0}^T Acc\_VNR}{\int_{t=0}^T All\_VNR}\\&=\lim_{T\rightarrow + \infty}\frac{\sum^T_{t=0}Acc\_VNR}{\sum^T_{t=0}All\_VNR},
\end{split}
\label{eq:26}
\end{align}
where Eq. \ref{eq:24} represents the \textit{long-term average revenue}, which is formally expressed as the average integral of the embedding revenues of all VNRs over time in the long-term range $[0, T]$. Eq. \ref{eq:25} represents the \textit{long-term average revenue-cost ratio}, which is formally expressed as the average integral of the embedding revenues over the embedding cost of all VNRs in the long-term range $[0, T]$. Eq. \ref{eq:26} represents the \textit{VNR acceptance ratio} in the VNE embedding process, which is formally expressed as the ratio of the successfully responded VNR to all the VNRs. It should be noted here that $Acc\_VNR$ indicates the number of successfully responded VNRs, which is expressed as:
\begin{equation}
    Acc\_VNR = \sum_{i=1}^{|VNR|} |VNR_i|,
\end{equation}
\begin{equation}
    |VNR_i| = \prod_{j=1}^{|N^{V_i}|}\zeta_{n^{P}\rightarrow n^{V_i}_j} \times
    \prod_{k=1}^{|L^{V_i}|}\xi_{l^{P}\rightarrow l^{V_i}_k}.
\end{equation}

\section{HFL-VNE Algorithm Design and Implementation}\label{sec:5}
In the following content, we will introduce the architecture design and implementation details of proposed HFL-VNE.

\subsection{Architecture Design}
Taking into account the poor performance of traditional VNE algorithms in dynamic and time-varying physical network environments, the distributed ML model of FL layout can better pay attention to the characteristics of each local physical domain and decrease the fragmentation of physical resources. Moreover, considering the weaknesses of weak privacy, high central pressure, and poor real-time performance caused by the traditional centralized training paradigm, FL can effectively alleviate such problems through multi-model collaborative optimization. Furthermore, in VNE, we utilize DRL as a learning-driven model. Compared with other ML models, DRL can better interact with the environment efficiently, and better explore and analyze the physical environment information. And, the distributed training paradigm of FL enables different DRL models to intelligently exchange parameters and collaboratively optimize without sharing data. In this way, each local DRL model can accurately grasp the global physical network. In addition, only weights or parameters are shared in the FL paradigm, which not only strengthens the confidentiality of data, but also relieves the pressure on computing and bandwidth.

Based on the above motivations, we deploy the FL training paradigm for multi-domain VNE, and deploy the DRL model in the local server and the global server respectively. Since multi-domain physical networks share the same feature space, we adopt the HFL paradigm and all DRL models adopt a similar structure. The specific architecture of HFL-VNE is displayed in Fig. \ref{fig:model}.

\subsection{Implementation Details}
In the proposed HFL-VNE, the iterative optimization process mainly includes three stages, as follows:

\subsubsection{State extraction} DRL agents need to efficiently interact with the environment, especially highly dynamic time-varying physical networks. It is necessary to monitor resource changes in the physical environment at all times and make appropriate resource scheduling strategies for the current VNR. This requires the agent to perceive the environment in real-time and extract environmental features as state input. Specifically, we extract the following important information: available CPU, the sum of bandwidth, and average distance.
\begin{equation}
    Sum\_bw(n^P_k)={\sum}_{\forall l^P_{jk}} A\_L(l_k^P),
    \label{eq:29}
\end{equation}
\begin{equation}
    Dis(n^P_k) = \frac{{\sum}_{\forall l^P_{jk}} \left\| n^P_k - n^P_j \right\|_2}{1+hops(l^P_{jk})},
    \label{eq:30}
\end{equation}
where taking the physical node $n^P_k$ as an example, its available CPU is shown in Eq. \ref{eq:16}. Eq. \ref{eq:29} implies the sum of the bandwidths of all links connecting $n^P_k$, which is inspired by the more available bandwidth of a link, the easier it is to be embedded. Eq. \ref{eq:30} implies the distance of $n^P_k$ to other nodes. It is inspired by the lower the hops and the shorter the distance, the higher the resource revenue.
Therefore, the extracted state matrix is shown in Eq.\ref{eq:31}, which will serve as the input of the intelligent agent of the DRL model.
\begin{equation}
\small
    State=\begin{pmatrix}
    A\_N(n^P_1) & Sum\_bw(n^P_1) & Dis(n^P_1)\\
    A\_N(n^P_2) & Sum\_bw(n^P_2) & Dis(n^P_2)\\
    \vdots & \vdots & \vdots\\
    A\_N(n^P_{|N^{P_i}|}) & Sum\_bw(n^P_{|N^{P_i}|}) & Dis(n^P_{|N^{P_i}|})\\
    \end{pmatrix}.
    \label{eq:31}
\end{equation}

\subsubsection{Agent Model of DRL}
The structure of the intelligent agent utilized in this work is displayed in Fig. \ref{fig:model}\textcolor{red}{(b)}, which is mainly a four-layer network architecture:
\begin{itemize}
    \item State extraction layer: This layer is mainly responsible for extracting the state of the physical domain in real-time and converting it into a state matrix as shown in Eq. \ref{eq:31}.
    \item Convolutional layer: This layer is responsible for the convolution operation on the state matrix with the support of the convolution kernel, and then obtains the available resources of each physical node and physical link. Furthermore, the convolution operation employed is mainly represented as:
    \begin{equation}
        Output(State) = Input(State)\cdot Kernel+ d,
    \end{equation}
    where $Kernel$ indicates the weight matrix of the convolution kernel, and $d$ indicates the deviation vector.
    \item Probability calculation layer: This layer employs the Softmax function to calculate the probability that the resources of physical nodes can be allocated.
    \begin{equation}
        Pro(Output[k]) = \frac{Output[k]}{\sum_{j=1}^{|N^{P_i}|}}Output[j].
    \end{equation}
    \item Filter layer: This layer filters the physical nodes and physical links that do not meet the conditions according to the embedding constraints shown in Section \ref{sec:filter}, and sorts the physical nodes in descending order according to the calculated probability. Finally, candidate physical nodes with priorities from high to low are obtained.
\end{itemize}

In HFL-VNE, Action is divided into two-stage decision-making actions. Specifically, node embedding is performed on the candidate physical nodes in order from high to low. Second, a search strategy--Breadth-First Search (BFS) is adopted for link embedding. The $LtaR2C$ adopted by the reward mechanism, under its guidance, the agent will optimize toward high-revenue and low-cost decision-making. If the value of the reward becomes larger, the current action is correct. Otherwise, the action needs to be adjusted.
\begin{equation}
    \small
    Maximize\left(Reward(V_i, t_s, t_e)\right) = LtaR2C (V_i, t_s, t_e).
\end{equation}
\subsection{Parameter Aggregation}
In the FL paradigm, the trained local model parameters of the local servers deployed in each physical domain will be uploaded to the global server. The latter will aggregate these parameters and performs iterative optimization. Finally, the new global parameter $\theta '$ is distributed to each local server. The local server updates the parameters and performs the next training. The above process is repeated until the HFL-VNE reaches the optimum, i.e., the objective function and the reward both converge.

In addition, for the global model, its DRL agent receives the federated state between the physical domains. For each physical domain $P_i$, its extracted state is $State_i(t)$, and the action taken by the agent of the local model is $Action_i(t)$. For the environment with $N$ physical domains, the federated state of the global model is $State_{all}(t)$ and the federated action is $Action_{all}(t)$. After the global model is iteratively optimized, new global parameters are obtained. At the same time, the federated reward $Reward_{all}(t)$ will be obtained. Further, the resources of each physical domain will be allocated, and the federated state will be transformed into the new state $State_{all}(t+1)$.
\begin{align}
\footnotesize
    &State_{all}(t) = \left\{State_1(t), \cdots, State_N(t)\right\},\\
&Action_{all}(t) = \left\{Action_1(t), \cdots, Action_N(t)\right\},\\
&Reward_{all}(t) = \left\{Reward_1(t),  \cdots, Reward_N(t)\right\},\\
&State_{all}(t+1) = \left\{State_1(t+1), \cdots, State_N(t+1)\right\}.
\end{align}
\section{Simulation Case Studies and Analysis}\label{sec:6}
In the following content, simulation cases and comparative experiments will be conducted to illustrate the effectiveness of HFL-VNE.
\subsection{Simulation Environment}
\begin{table}[!htbp]
\caption{The detailed configurations of the physical network and virtual networks}
\label{tab:2}
\centering
\renewcommand\arraystretch{1.3}
\begin{tabular}{|l|l|}
\hline
\textbf{Physical Network} & \textbf{Configuration} \\ \hline
Number of physical domains         & 4                      \\ 
Number of nodes per domain      & 25                     \\ 
Number of physical links             & 600                    \\ 
CPU of nodes              & {[}50,100{]}           \\ 
BW of links               & {[}50,100{]}           \\ \hline \hline
\textbf{Virtual Network} & \textbf{Configuration} \\ \hline
Number of VNRs & 2000\\
Training set size & 1000\\
Testing set size & 1000 \\
CPU of virtual nodes & [1,50]\\
BW of virtual links& [1,50]\\ \hline
\end{tabular}
\end{table}
\begin{figure*}[!htbp]
    \centering
    \subfigure[$LtaR$ on training]{
        \includegraphics[width=0.3\linewidth]{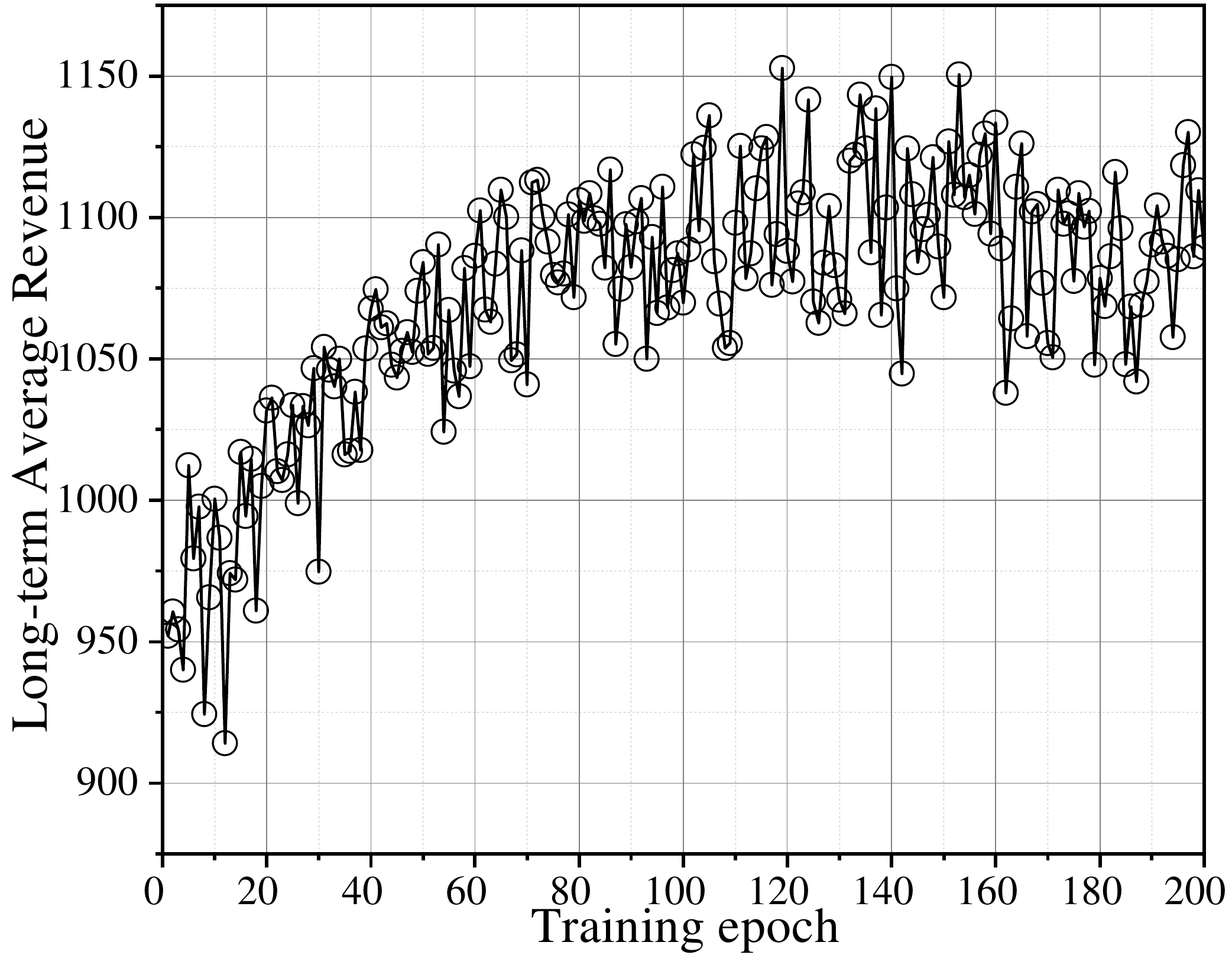}
        \label{fig: lr}
    }
    \subfigure[$LtaR2C$ on training]{
        \includegraphics[width=0.3\linewidth]{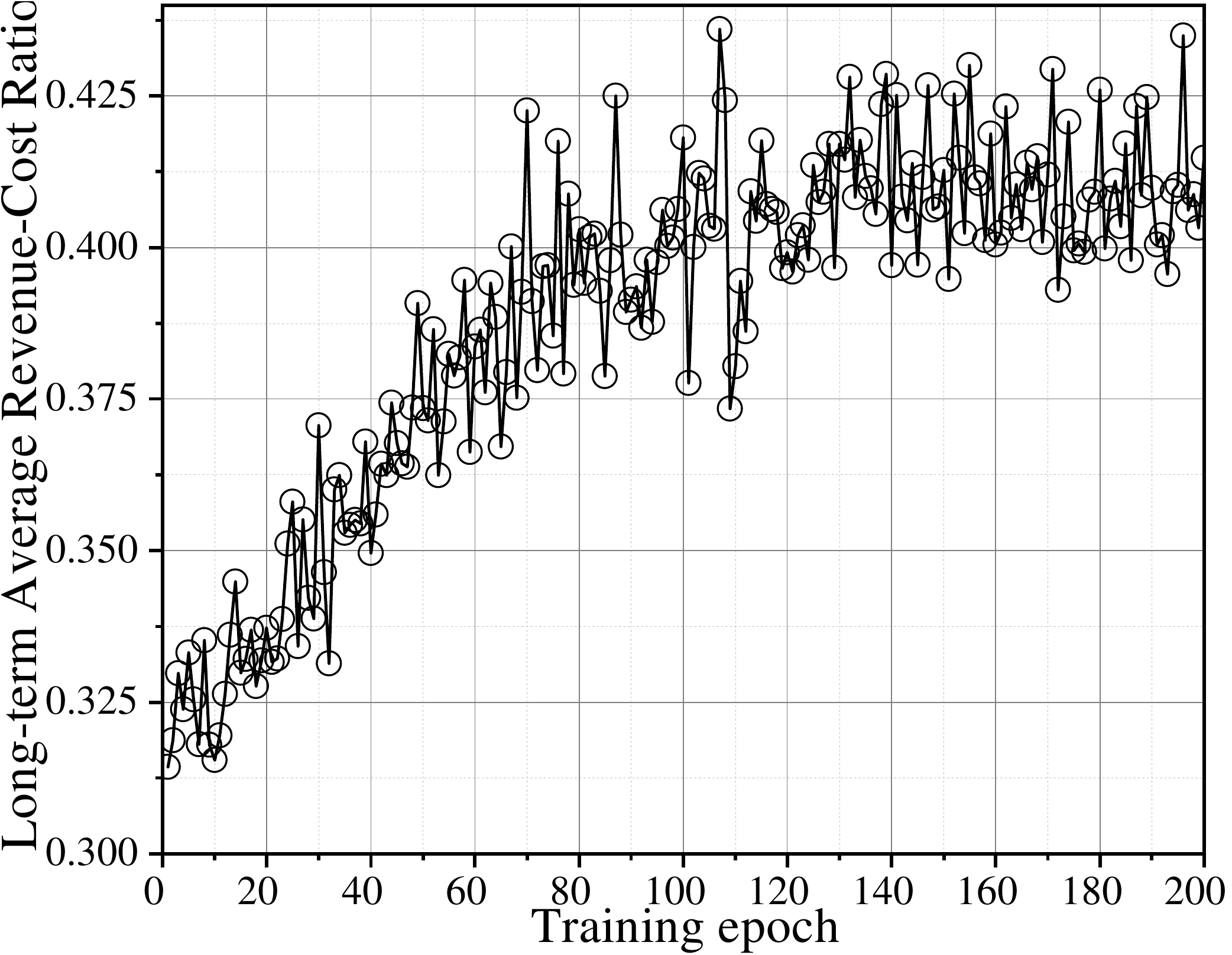}
        \label{fig: rc}
    }
    \subfigure[$Acc$ on training]{
        \includegraphics[width=0.3\linewidth]{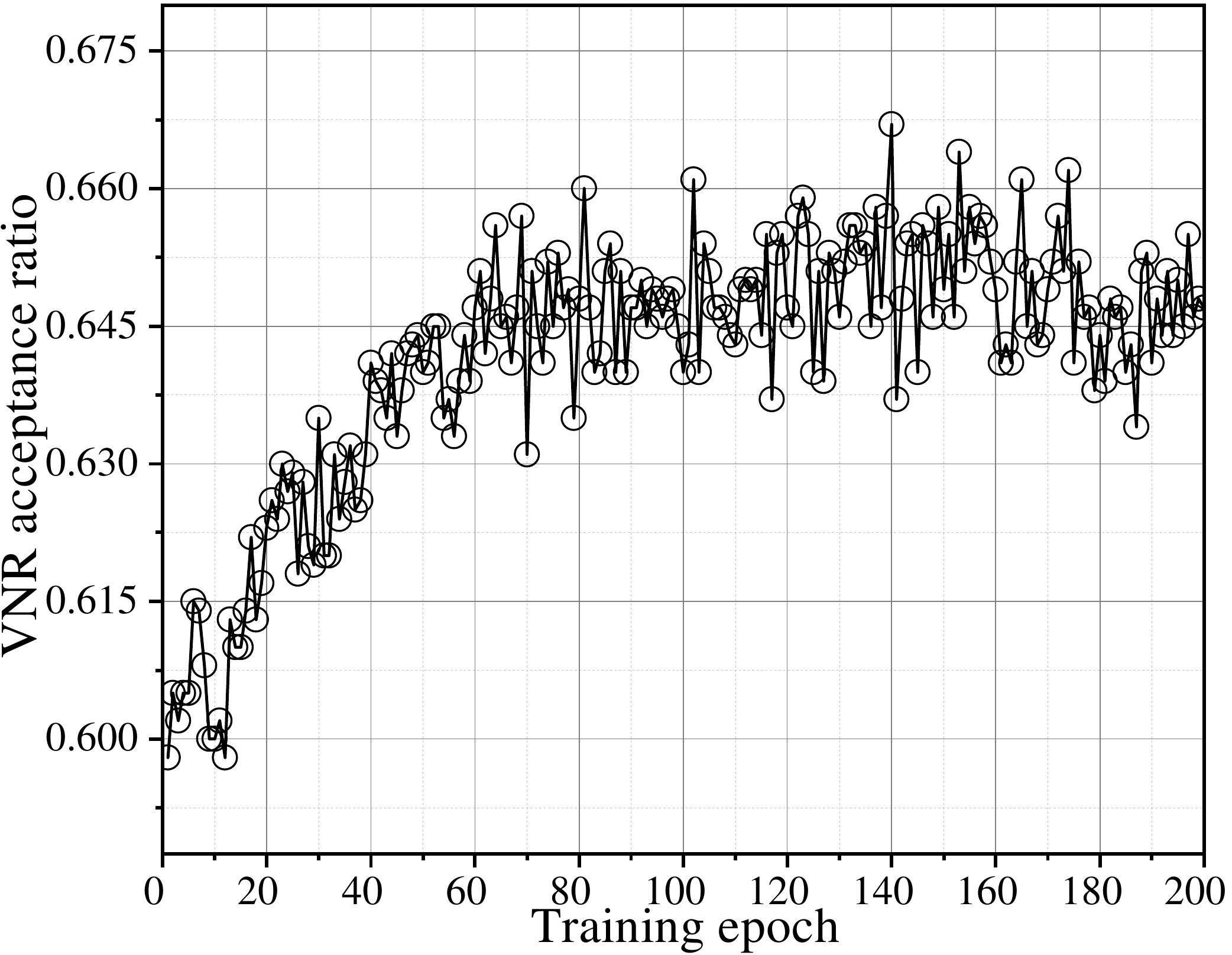}
        \label{fig: acc}
    }
    \caption{Performance of HFL-VNE in training phases}
    \label{fig:case}
\end{figure*}
The simulation experiment is completed by \textit{Pycharm} and using \textit{python 3.8.3}. The physical network is generated using the \textit{GT-ITM} tool and recorded in a \textit{.txt} file. Specifically, we generate a physical network of 4 domains, each with 25 physical nodes. There are a total of 100 physical nodes and 600 physical links. Furthermore, 2000 VNRs are randomly generated and recorded in \textit{.txt} files. In each VNR, virtual links are randomly generated between virtual nodes with a probability of 50\%. To simulate the dynamic continuity of the VNR, it arrives sequentially in the time order of the Poisson distribution. The detailed configurations of the physical network and virtual networks are recorded in Table \ref{tab:2}.

\subsection{Training Performance}
To illustrate the flexibility and intelligence of the proposed HFL-VNE in the face of multi-domain dynamic environments, it is necessary to observe the trend of the performance indicators to scientifically verify that it is reasonably allocating the resources. In the training phase of HFL-VNE, we observe the performance of each evaluation indicator. 
The indicators variations of the HFL-VNE training process are displayed in Fig. \ref{fig:case}.
From the visualized result trend, it can be found that the performance of each indicator is relatively poor at the beginning. This is since all agents have just been activated in an unfamiliar environment and are not suitable for it. As the training epoch increases, it can quickly adapt to the environment and fit the underlying characteristics of the physical network, and stimulate it to continuously allocate high-revenue and high-efficient decision-making through the reward mechanism, which makes various indicators rapidly improve and converge. Therefore, the proposed HFL-VNE is effective in the training process of the multi-domain physical network.
\begin{figure}
    \centering
    \includegraphics[width=0.9\linewidth]{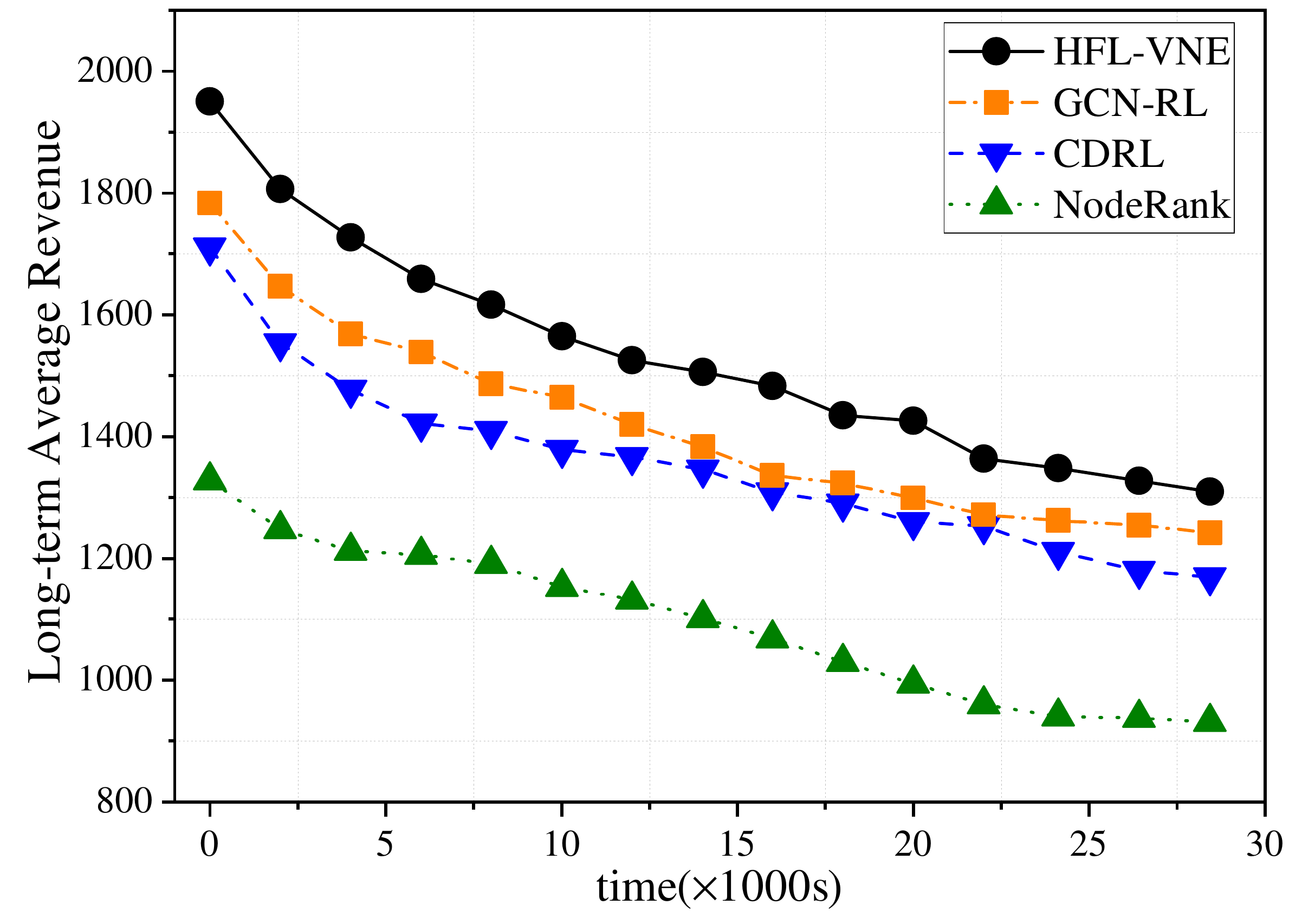}
    \caption{Performance comparison of HFL-VNE and benchmarks on $LtaR$.}
    \label{fig:comlr}
\end{figure}
\begin{figure}
    \centering
    \includegraphics[width=0.9\linewidth]{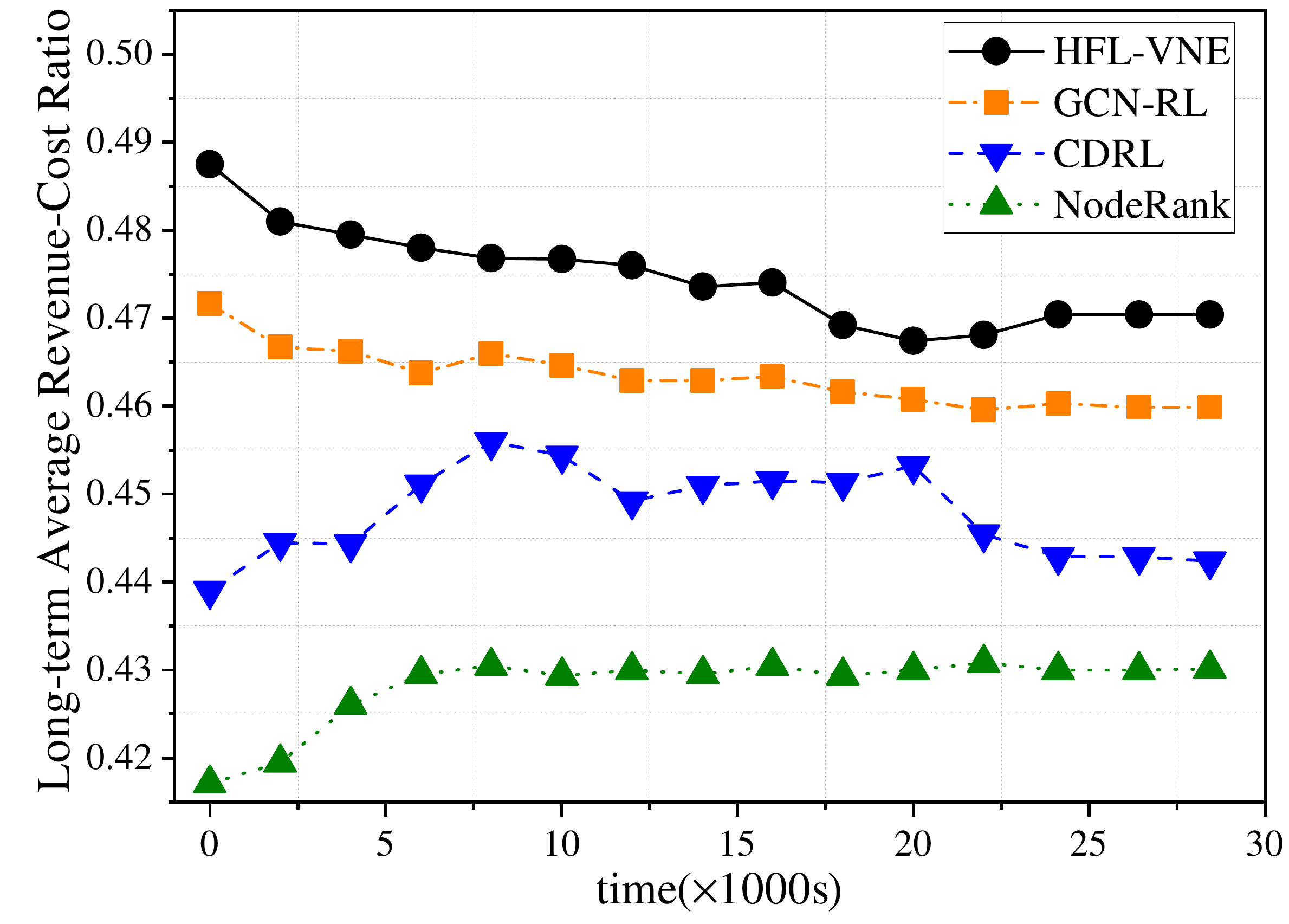}
    \caption{Performance comparison of HFL-VNE and benchmarks on $LtaR2C$.}
    \label{fig:comlrc}
\end{figure}
\begin{figure}
    \centering
    \includegraphics[width=0.9\linewidth]{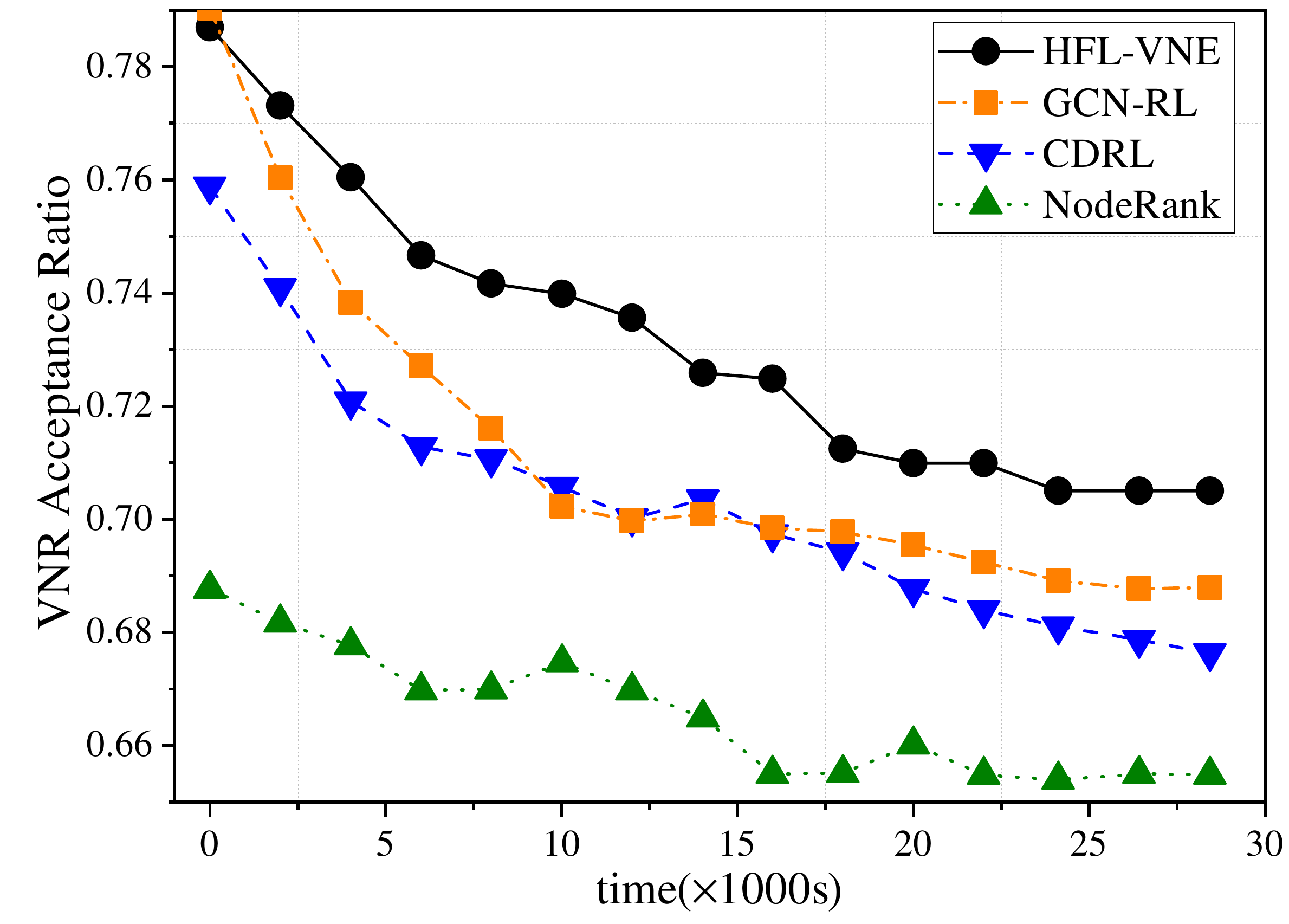}
    \caption{Performance comparison of HFL-VNE and benchmarks on $Acc$.}
    \label{fig:acc}
\end{figure}
\begin{figure}
    \centering
    \includegraphics[width=0.8\linewidth]{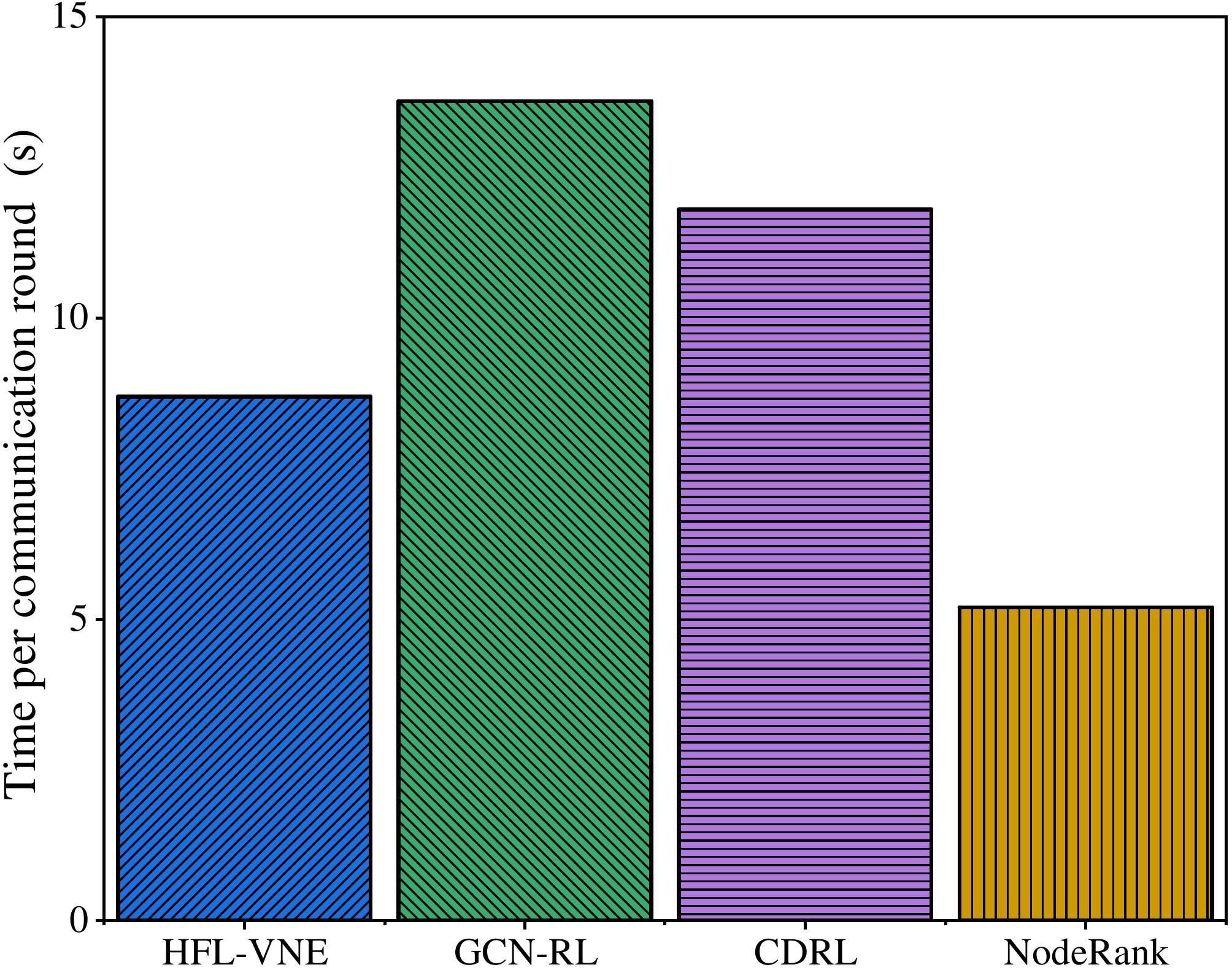}
    \caption{Time comparison of each communication round.}
    \label{fig:time}
\end{figure}
\subsection{Comparative Experiment}
For comparative experiments, we selected heuristic-based, RL-based and DRL-based algorithms as benchmarks, namely NodeRank \cite{cheng2011virtual}, CDRL \cite{yao2020continuous} and GCN-RL \cite{9475485}, respectively. On the three evaluation indicators, the performance of HFL-VNE and each benchmark is presented in Figs. \ref{fig:comlr}, \ref{fig:comlrc}, and \ref{fig:acc}, respectively.
It can be found that the NodeRank based on the heuristic strategy preferentially selects nodes with a heavyweight at the beginning, so the initial performance is reasonable. However, from the perspective of long-term performance, it is difficult to effectively characterize dynamic multi-domain networks and thus perform poorly. Combined with RL, CDRL increases the ability to interact with the environment, making model decisions perform well. On this basis, GCN-RL further uses GCN to replace the traditional Q-table, which enhances the perception ability of the model and further enables the VNE strategy to better perceive the network environment.
Furthermore, HFL-VNE outperforms both CDRL and GCN-RL in the long run. This is mainly because we deploy local DRL models in various physical domains, which can better focus on the properties of each local physical domain. In dynamic and heterogeneous network modeling, HFL-VNE performs better, thereby better reducing the fragmentation of physical resources.

In addition, during the test, the server continuously responded to the VNR, which made the available resources of the physical network tend to decrease. Correspondingly, the number of VNRs embedded in the next step will be reduced. According to Eq. \ref{eq:24}, \ref{eq:25}, and \ref{eq:26}, it can be seen that the $LtaR$ and the $Acc$ will show a decreasing trend. The $LtaR2C$ is not only related to benefits but also costs, so this indicator shows a trend of small fluctuations. The comprehensive results well validate our theoretical derivation.

Finally, in Fig. \ref{fig:time}, we record the time spent by HFL-VNE with each benchmark in each communication round. NodeRank takes less time because heuristic search uses a simple greedy search strategy. However, its performance is poor. CDRL and GCN-RL only deploy one model and adopt a centralized training strategy, which increases the complexity of the algorithm and increases the pressure on storage and computation. Therefore, both are more time-consuming than HFL-VNE. Combined with the distributed FL training model, HFL-VNE is jointly optimized on the premise of only uploading parameters, which significantly reduces the time-consuming of resource embedding. Therefore, HFL-VNE has better real-time performance than other algorithms. To sum up, HFL-VNE outperforms other VNE algorithms in all aspects.
\section{Conclusion}\label{sec:7}
To solve the virtualization issue of multi-domain physical networks, this work is the first to propose a virtual network embedding strategy based on horizontal federated learning (HFL-VNE). Specifically, we utilize the distributed architecture of FL to deploy a local server in each physical domain, which is responsible for the perception and decision-making of the local physical domain. Furthermore, we deploy a global server on the overall architecture for aggregating parameters to share training. Also, we deploy a deep reinforcement learning model in each server, which is employed to dynamically tune and optimize the VNE policy. Finally, we effectively prove by simulation that HFL-VNE is more suitable for modeling dynamic and heterogeneous physical networks, and HFL-VNE has higher privacy and real-time performance.
\bibliographystyle{IEEEtran}
\bibliography{Main}
\begin{IEEEbiography}[{\includegraphics[width=1in,height=1.25in,clip,keepaspectratio]{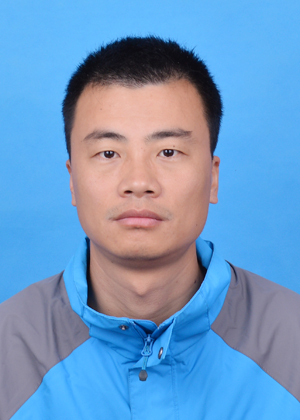}}]{Peiying Zhang} [M'21] is currently an Associate Professor with the College of Computer Science and Technology, China University of Petroleum (East China). He received his Ph.D. in the School of Information and Communication Engineering at University of Beijing University of Posts and Telecommunications in 2019. He has published multiple IEEE/ACM Trans./Journal/Magazine papers since 2016, such as IEEE TII, IEEE T-ITS, IEEE TVT, IEEE TNSE, IEEE TNSM, IEEE TETC, IEEE Network, IEEE Access, IEEE IoT-J, ACM TALLIP, COMPUT COMMUN, IEEE COMMUN MAG, and etc. He served as the Technical Program Committee of IEEE ICC’22, DPPR 2021, ISCIT 2016, ISCIT 2017, ISCIT 2018, ISCIT 2019, Globecom 2021, Globecom 2019, COMNETSAT 2020, SoftIoT 2021, IWCMC-Satellite 2019, IWCMC-Satellite 2020, and IWCMC-Satellite 2022. His research interests include semantic computing, future internet architecture, network virtualization, and artificial intelligence for networking.  \end{IEEEbiography}

\begin{IEEEbiography}[{\includegraphics[width=1in,height=1.25in,clip,keepaspectratio]{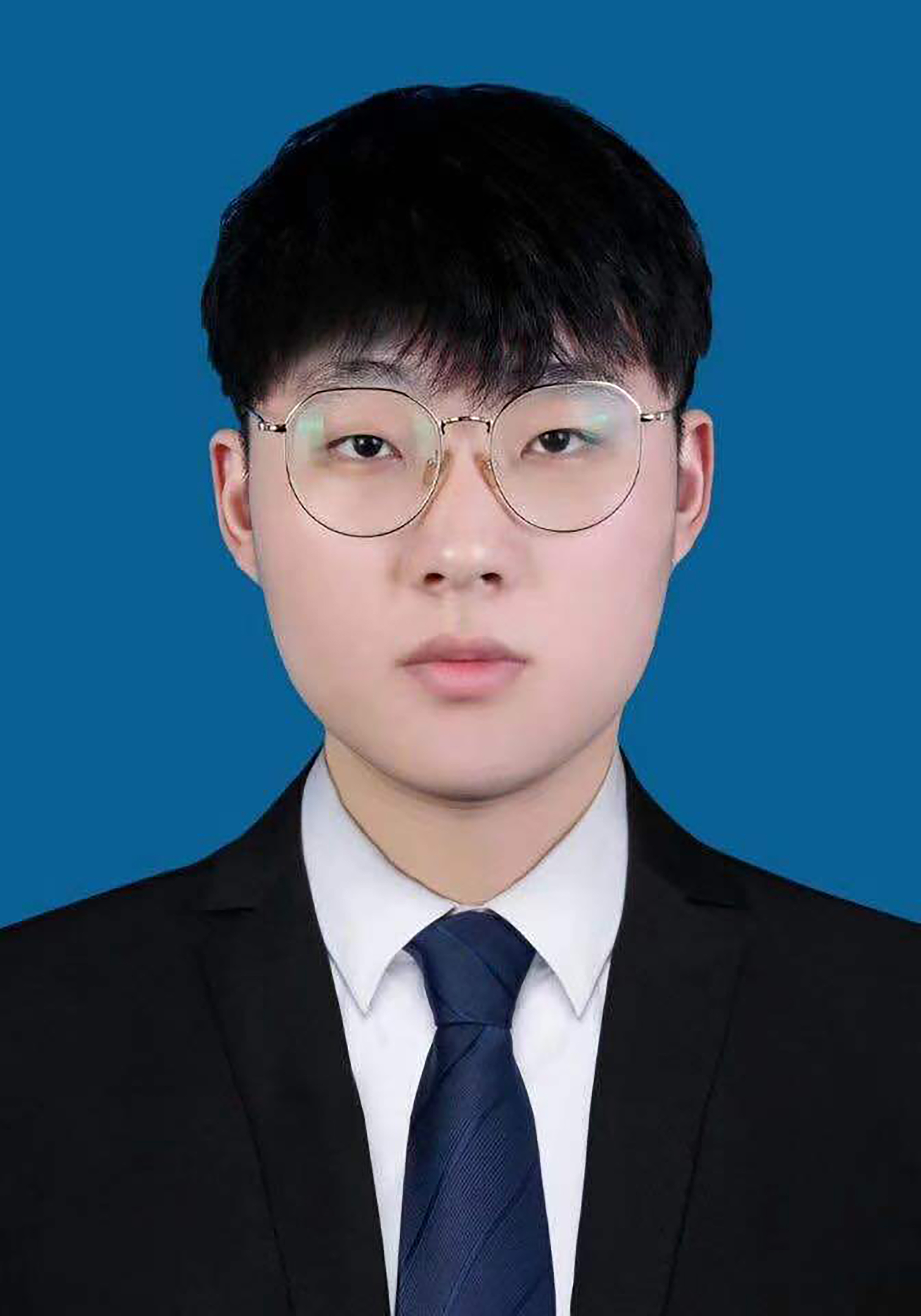}}]{Ning Chen} [S'22]
is currently studying for a master's degree in the the College of Computer Science and Technology, China University of Petroleum (East China). His research interests include deep learning, network virtualization, and future network architecture.
 \end{IEEEbiography}
 
 \begin{IEEEbiography}[{\includegraphics[width=1in,height=1.25in,clip,keepaspectratio]{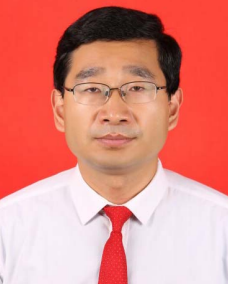}}]{Shibao Li}
received the B.S. and M.S. degrees
in computer science from the China University
of Petroleum, Qingdao, China, in 2002 and 2009,
respectively. He is currently a professor of college of oceanography and space informatics, China University of Petroleum, Qingdao, 26658, China. His main research interests are mobile computing, wireless big data, broadband wireless communication, and intelligent information processing.
 \end{IEEEbiography}
 
\begin{IEEEbiography}[{\includegraphics[width=1in,height=1.25in,clip,keepaspectratio]{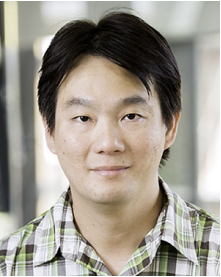}}]{Kim-Kwang Raymond Choo} [SM'15] received the Ph.D. degree in information security from the Queensland University of Technology, Australia, in 2006. He currently holds the Cloud Technology Endowed Professorship at the University of Texas at San Antonio (UTSA). He is the founding Co-Editor-in-Chief of ACM Distributed
Ledger Technologies: Research \& Practice (commencing June 2021), founding Chair of IEEE Technology and Engineering Management Society’s Technical Committee on Blockchain and Distributed Ledger Technologies. He is an ACM Distinguished Speaker and IEEE Computer Society Distinguished Visitor during 2021–2023, and included in Web of Science’s Highly Cited Researcher in the field of Cross-Field2020. He has been cited more than 24448 times (H-index = 92).
In 2016, he was named the Cybersecurity Educator of the Year-APAC (Cybersecurity Excellence Awards are produced in cooperation with the Information Security Community on LinkedIn).
His research has been funded by the National Science Foundation, NASA, CPS Energy, LGS Innovations, the Texas National Security Network Excellence Fund, the Australian Government National Drug Law Enforcement Research Fund, the Australian Government Cooperative Research Centre for Data to Decision, the auDA Foundation, the Government of South Australia, BAE Systems strategic, the Australasian Institute of Judicial Administration Incorporated, and the Australian Research Council. He is a fellow of the Australian Computer Society and the Co-Chair of the IEEE Multimedia Communications Technical Committee’s Digital Rights Management for Multimedia Interest Group. In 2015, he and his team won the Digital Forensics Research Challenge organized by the Germany’s University of Erlangen–Nuremberg. He was a recipient of the 2019 IEEE Technical Committee on Scalable Computing (TCSC) Award for Excellence in Scalable Computing (Middle Career Researcher), the 2018 UTSA College of Business Col. Jean Piccione and Lt. Col. Philip Piccione Endowed Research Award for Tenured Faculty, an Outstanding Associate Editor of 2018 for IEEE ACCESS, the British Computer Society’s 2019 Wilkes Award Runner-up, the 2019 EURASIP Journal on Wireless Communications and Networking (JWCN) Best Paper Award, the Korea Information Processing Society’s Journal of Information Processing Systems (JIPS) Survey Paper Award (Gold) 2019, the IEEE Blockchain 2019 Outstanding Paper Award, the Inscrypt 2019 Best Student Paper Award, the IEEE TrustCom 2018 Best Paper Award, the ESORICS 2015 Best Research Paper Award, the 2014 Highly Commended Award by the Australia New Zealand Policing Advisory Agency, the Fulbright Scholarship in 2009, the 2008 Australia Day Achievement Medallion, and the British Computer Society’s Wilkes Award in 2008.
 \end{IEEEbiography}
 \begin{IEEEbiography}[{\includegraphics[width=1in,height=1.25in,clip,keepaspectratio]{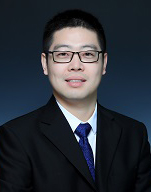}}]{Chunxiao Jiang} [S’09-M’13-SM’15] is an associate professor in School of Information Science and Technology, Tsinghua University. He received the B.S. degree in information engineering from Beihang University, Beijing in 2008 and the Ph.D. degree in electronic engineering from Tsinghua University, Beijing in 2013, both with the highest honors. From 2011 to 2012 (as a Joint Ph.D) and
2013 to 2016 (as a Postdoc), he was in the Department of Electrical and Computer Engineering at University of Maryland College Park under the
supervision of Prof. K. J. Ray Liu. His research interests include application
of game theory, optimization, and statistical theories to communication,
networking, and resource allocation problems, in particular space networks
and heterogeneous networks. 

Dr. Jiang has served as an Editor of IEEE Transactions on Communications, IEEE Internet of Things Journal, IEEE Wireless Communications, IEEE Network, IEEE Communications Letters, and a Guest Editor of IEEE Communications Magazine, IEEE Transactions on Network Science and Engineering and IEEE Transactions on Cognitive Communications and Networking. He has also served as a member of the technical program committee as well as the Symposium Chair for a number of international conferences, including IEEE Globecom 2021 Symposium Chair,
IEEE CNS 2020 Publication Chair, IEEE WCSP 2019 Symposium Chair, IEEE ICC 2018 Symposium Co-Chair, IWCMC 2020/19/18 Symposium Chair, WiMob 2018 Publicity Chair, ICCC 2018 Workshop Co-Chair, and IEEE ICC 2017 Workshop Co-Chair. Dr. Jiang is the recipient of the Best Paper Award from IEEE GLOBECOM in 2013, the Best Student Paper Award from IEEE GlobalSIP in 2015, IEEE Communications Society Young Author
Best Paper Award in 2017, the Best Paper Award IWCMC in 2017, IEEE
ComSoc TC Best Journal Paper Award of the IEEE ComSoc TC on Green
Communications \& Computing 2018, IEEE ComSoc TC Best Journal Paper
Award of the IEEE ComSoc TC on Communications Systems Integration
and Modeling 2018, the Best Paper Award from ICC 2019, IEEE VTS
Early Career Award 2020, IEEE ComSoc Asia-Pacific Best Young Researcher
Award 2020, and IEEE VTS Distinguished Lecturer 2021. He received the
Chinese National Second Prize in Technical Inventions Award in 2018 and
Natural Science Foundation of China Excellent Young Scientists Fund Award
in 2019. He is a Senior Member of IEEE and a Fellow of IET.

 \end{IEEEbiography}
\end{document}